\documentclass[10pt]{revtex4}    
\usepackage{amssymb}
\usepackage{latexsym}
\usepackage{epsfig}
\begin{document}

\title{Interacting agegraphic quintessence dark energy in non-flat universe}

\author{A. Sheykhi$^{a,b}$\footnote{sheykhi@mail.uk.ac.ir}, A. Bagheri$^{a}$ and
M.M. Yazdanpanah$^{a}$\footnote{myazdan@mail.uk.ac.ir}}

\address{$^a$Department of Physics, Shahid Bahonar University, P.O. Box 76175, Kerman, Iran\\
         $^b$Research Institute for Astronomy and Astrophysics of Maragha (RIAAM), Maragha,
         Iran}

\begin{abstract}
We suggest a correspondence between interacting agegraphic dark
energy models and the quintessence scalar field in a non-flat
universe. We demonstrate that the agegraphic evolution of the
universe can be described completely by a single quintessence
scalar field. Then, we reconstruct the potential of the
interacting agegraphic quintessence dark energy  as well as the
dynamics of the scalar field according to the evolution of the
agegraphic dark energy.
\end{abstract}
\maketitle
\section{Introduction\label{Int}}
It is a general belief that our universe is currently experiencing
a phase of accelerated expansion \cite{Rie}. Missing energy
component with negative pressure which is responsible for this
expansion constitute a major puzzle of modern cosmology. Despite
the theoretical difficulties in understanding dark energy,
independent observational evidence for its existence is
impressively robust. Among the various candidates to explain the
accelerated expansion, the cosmological constant with $w=-1$ is
located at a central position.  Though, it suffers the so-called
{fine-tuning} and {cosmic coincidence} problems. In quintessence
\cite{QUINT} and  Chaplygin gas \cite{KMP} $w$ always stays bigger
than $-1$. The phantom models of dark energy have $w<-1$
\cite{PHANT}. However, following the more accurate data analysis,
a more dramatic result appears showing that the time varying dark
energy gives a better fit than a cosmological constant and in
particular, $w$ can cross $-1$ from above to below \cite{Alam}.

An interesting attempt for probing the nature of dark energy
within the framework of quantum gravity is the holographic dark
energy. This proposal, that arose a lot of enthusiasm recently
\cite{Coh,Li,Huang,Hsu,HDE,Setare1}, is motivated from the
holographic hypothesis \cite{Suss1} and has been tested and
constrained by various astronomical observations \cite{Xin}.
However there are some difficulties in holographic dark energy
model. Choosing the event horizon of the universe as the length
scale, the holographic dark energy gives the observation value of
dark energy in the universe and can drive the universe to an
accelerated expansion phase. But an obvious drawback concerning
causality appears in this proposal. Event horizon is a global
concept of spacetime; existence of event horizon of the universe
depends on future evolution of the universe; and event horizon
exists only for universe with forever accelerated expansion. In
addition, more recently, it has been argued that this proposal
might be in contradiction to the age of some old high redshift
objects, unless a lower Hubble parameter is considered
\cite{Wei0}. Another proposal to probe the nature of dark energy
within the framework of quantum gravity is a so-called agegraphic
dark energy (ADE). This model is based on the uncertainty relation
of quantum mechanics together with the gravitational effect in
general relativity. Following the line of quantum fluctuations of
spacetime, Karolyhazy et al. \cite{Kar1} argued that the distance
$t$ in Minkowski spacetime cannot be known to a better accuracy
than $\delta{t}=\beta t_{p}^{2/3}t^{1/3}$ where $\beta$ is a
dimensionless constant of order unity. Based on Karolyhazy
relation, Maziashvili discussed that the energy density of metric
fluctuations of the Minkowski spacetime is given by \cite{Maz}
\begin{equation}\label{rho0}
\rho_{D} \sim \frac{1}{t_{p}^2 t^2} \sim \frac{m^2_p}{t^2},
\end{equation}
where $t_{p}$ is the reduced Planck time. We use the units $c
=\hbar=k_b = 1$ throughout this work. Therefore one has $l_p = t_p
= 1/m_p$ with $l_p$ and $m_p$ are the reduced Planck length and
mass, respectively. The agegraphic dark energy model assumes that
the observed dark energy comes from the spacetime and matter field
fluctuations in the universe \cite{Cai1,Wei2,Wei1}. Since in
agegraphic dark energy model the age of the universe is chosen as
the length measure, instead of the horizon distance, the causality
problem in the holographic dark energy is avoided. The agegraphic
models of dark energy  have been examined and constrained by
various astronomical observations \cite{age,shey1,Setare2}.
Although going along a fundamental theory such as quantum gravity
may provide a hopeful way towards understanding the nature of dark
energy, it is hard to believe that the physical foundation of
agegraphic dark energy is convincing enough. Indeed, it is fair to
say that almost all dynamical dark energy models are settled at
the phenomenological level, neither holographic dark energy model
nor agegraphic dark energy model is exception. Though, under such
circumstances, the models of holographic and agegraphic dark
energy, to some extent, still have some advantage comparing to
other dynamical dark energy models because at least they originate
from some fundamental principles in quantum gravity. We thus may
as well view that this class of models possesses some features of
an underlying theory of dark energy.

On the other hand, the scalar field model is an effective
description of an underlying theory of dark energy. Scalar fields
naturally arise in particle physics including supersymmetric field
theories and string/M theory. Therefore, scalar field is expected
to reveal the dynamical mechanism and the nature of dark energy.
However, although fundamental theories such as string/M theory do
provide a number of possible candidates for scalar fields, they do
not uniquely predict its potential $V(\phi)$. Therefore it becomes
meaningful to reconstruct $V(\phi)$ from some dark energy models
possessing some significant features of the quantum gravity
theory, such as holographic and agegraphic dark energy model. The
investigations on the reconstruction of the potential $V(\phi)$ in
the framework of holographic dark energy have been carried out in
\cite{Zhang}. In the absence of the interaction between agegraphic
dark energy and dark matter, the quintessence reconstruction of
the agegraphic dark energy models have been established
\cite{ageQ}.

In this paper we intend to generalize the study to the case where
both components- the pressureless dark matter and the agegraphic
dark energy- do not conserve separately but interact with each
other. Given the unknown nature of both dark matter and dark
energy there is nothing in principle against their mutual
interaction and it seems very special that these two major
components in the universe are entirely independent
\cite{Setare3,wang1,shey2}. We shall establish a correspondence
between the interacting agegraphic dark energy scenarios and the
quintessence scalar field. We suggest the agegraphic description
of the quintessence dark energy in a universe with spacial
curvature and reconstruct the potential and the dynamics of the
quintessence scalar field which describe the quintessence
cosmology.

This paper is outlined as follows. In the next section we
demonstrate a correspondence between the original agegraphic and
quintessence dark energy model. In section \ref{NEW}, we establish
the correspondence between the new model of interacting agegraphic
dark energy and the quintessence dark energy. The last section is
devoted to conclusions.
\section{Quintessence reconstruction of ORIGINAL ADE  \label{ORI}}
We assume the agegraphic quintessence dark energy is accommodated
in the Friedmann-Robertson-Walker (FRW) universe which is
described by the line element
\begin{eqnarray}
 ds^2=dt^2-a^2(t)\left(\frac{dr^2}{1-kr^2}+r^2d\Omega^2\right),\label{metric}
 \end{eqnarray}
where $a(t)$ is the scale factor, and $k$ is the curvature
parameter with $k = -1, 0, 1$ corresponding to open, flat, and
closed universes, respectively. A closed universe with a small
positive curvature ($\Omega_k\simeq0.01$) is compatible with
observations \cite{spe}. The corresponding Friedmann equation
takes the form
\begin{eqnarray}\label{Fried}
H^2+\frac{k}{a^2}=\frac{1}{3m_p^2} \left( \rho_m+\rho_D \right).
\end{eqnarray}
We introduce, as usual, the fractional energy densities such as
\begin{eqnarray}\label{Omega}
\Omega_m=\frac{\rho_m}{3m_p^2H^2}, \hspace{0.5cm}
\Omega_D=\frac{\rho_D}{3m_p^2H^2},\hspace{0.5cm}
\Omega_k=\frac{k}{H^2 a^2},
\end{eqnarray}
thus, the Friedmann equation can be written
\begin{eqnarray}\label{Fried2}
\Omega_m+\Omega_D=1+\Omega_k.
\end{eqnarray}
We adopt the viewpoint that the scalar field models of dark energy
are effective theories of an underlying theory of dark energy. The
energy density and pressure for the quintessence scalar field can
be written as
\begin{eqnarray}\label{rhophi}
\rho_\phi=\frac{1}{2}\dot{\phi}^2+V(\phi),\\
p_\phi=\frac{1}{2}\dot{\phi}^2-V(\phi), \label{pphi}
\end{eqnarray}
Then, we can easily obtain the scalar potential and the kinetic
energy term as
\begin{eqnarray}\label{vphi}
&&V(\phi)=\frac{1-w_D}{2}\rho_{\phi},\\
&&\dot{\phi}^2=(1+w_D)\rho_\phi. \label{ddotphi}
\end{eqnarray}
Now we are focussing on the reconstruction of the original
agegraphic quintessence model of dark energy. The original
agegraphic dark energy density has the form (\ref{rho0}) where $t$
is chosen to be the age of the universe
\begin{equation}
T=\int_0^a{\frac{da}{Ha}},
\end{equation}
Thus, the energy density of the original agegraphic dark energy is
given by \cite{Cai1}
\begin{equation}\label{rho1}
\rho_{D}= \frac{3n^2 m_{p}^2}{T^2},
\end{equation}
where the numerical factor $3n^2$ is introduced to parameterize
some uncertainties, such as the species of quantum fields in the
universe, the effect of curved space-time (since the energy
density is derived for Minkowski space-time), and so on. The dark
energy density (\ref{rho1}) has the same form as the holographic
dark energy, but  the length measure is chosen to be the age of
the universe instead of the horizon radius of the universe. Thus
the causality problem in the holographic dark energy is avoided.
Combining Eqs. (\ref{Omega}) and (\ref{rho1}), we get
\begin{eqnarray}\label{Omegaq}
\Omega_D=\frac{n^2}{H^2T^2}.
\end{eqnarray}
The total energy density is $\rho=\rho_{m}+\rho_{D}$, where
$\rho_{m}$ and $\rho_{D}$ are the energy density of dark matter
and dark energy, respectively. The total energy density satisfies
a conservation law
\begin{equation}\label{cons}
\dot{\rho}+3H(\rho+p)=0.
\end{equation}
However, since we consider the interaction between dark matter and
dark energy, $\rho_{m}$ and $\rho_{D}$ do not conserve separately;
they must rather enter the energy balances
\begin{eqnarray}
&&\dot{\rho}_m+3H\rho_m=Q, \label{consm}
\\&& \dot{\rho}_D+3H\rho_D(1+w_D)=-Q.\label{consq}
\end{eqnarray}
Here $w_D$ is the equation of state parameter of agegraphic dark
energy and $Q$ denotes the interaction term and can be taken as $Q
=3b^2 H\rho$  with $b^2$  being a coupling constant. This
expression for the interaction term was first introduced in the
study of the suitable coupling between a quintessence scalar field
and a pressureless cold dark matter field \cite{Ame,Zim}. In the
context of holographic dark energy model, this form of interaction
was derived from the choice of Hubble scale as the IR cutoff
\cite{Pav1}. Taking the  derivative with respect to the cosmic
time of Eq. (\ref{rho1})  and using Eq. (\ref{Omegaq}) we get
\begin{eqnarray}\label{rhodot}
\dot{\rho}_D=-2H\frac{\sqrt{\Omega_D}}{n}\rho_D.
\end{eqnarray}
Inserting this relation into Eq. (\ref{consq}), we obtain the
equation of state parameter of the original agegraphic dark energy
\begin{eqnarray}\label{wq}
w_D=-1+\frac{2}{3n}\sqrt{\Omega_D}-\frac{b^2}{\Omega_D}
(1+\Omega_k).
\end{eqnarray}
\begin{figure}[htp]
\begin{center}
\includegraphics[width=8cm]{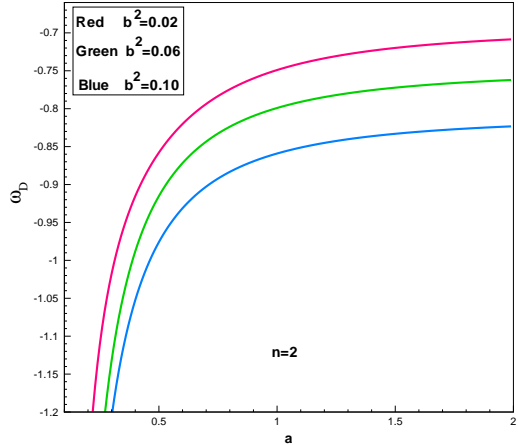}
\caption{The evolution of $w_D$ for original ADE with different
interacting parameter $b^2$. Here we take $\Omega_{D0}=0.72$ and
$\Omega_{k}=0.01.$ }\label{fig1}
\end{center}
\end{figure}
\begin{figure}[htp]
\begin{center}
\includegraphics[width=8cm]{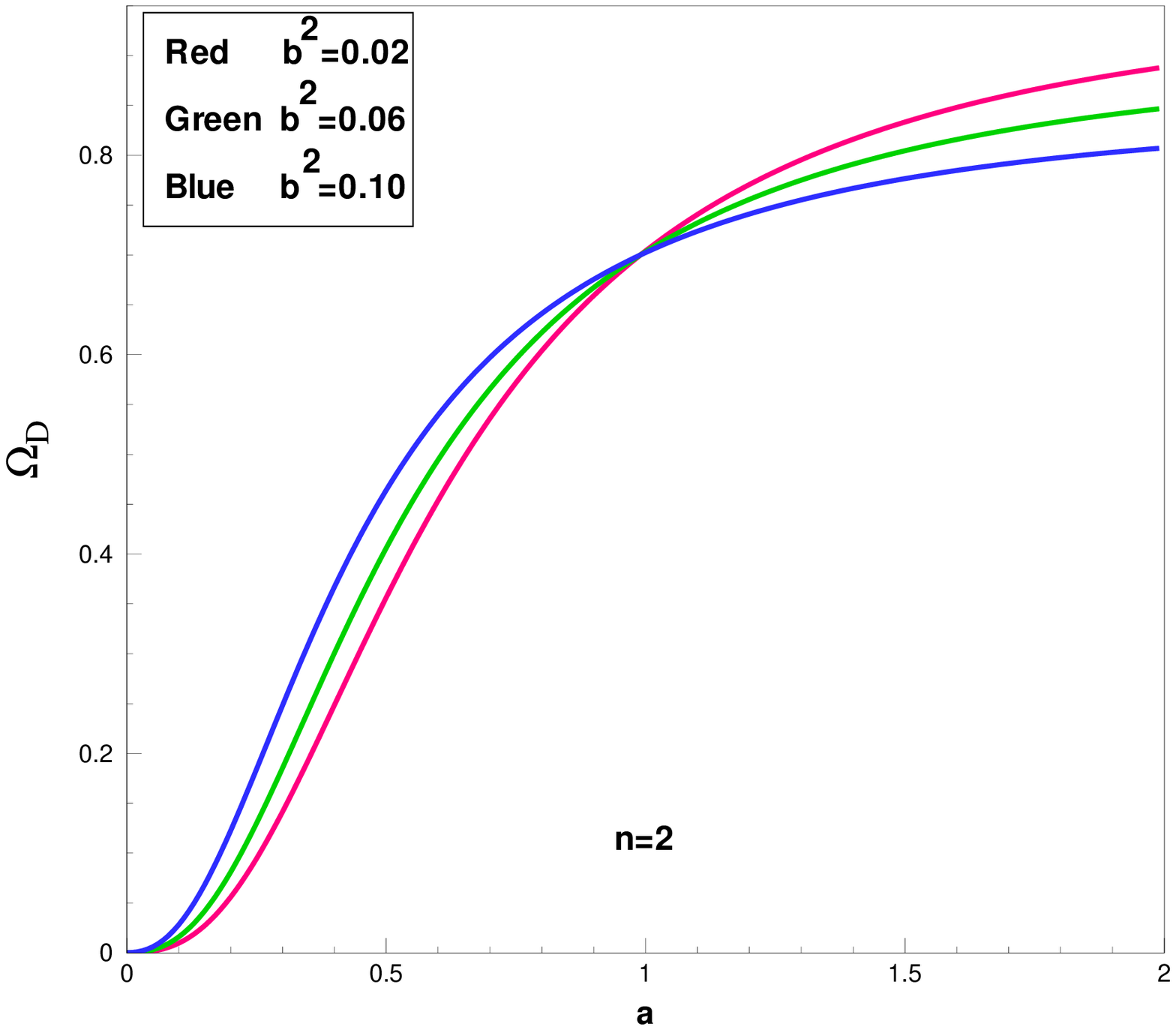}
\caption{The evolution of $\Omega_{D}$ for original ADE with
different interacting parameter $b^2$. Here we take
$\Omega_{D0}=0.72$ and $\Omega_{k}=0.01$.}\label{fig2}
\end{center}
\end{figure}
\begin{figure}[htp]
\begin{center}
\includegraphics[width=8cm]{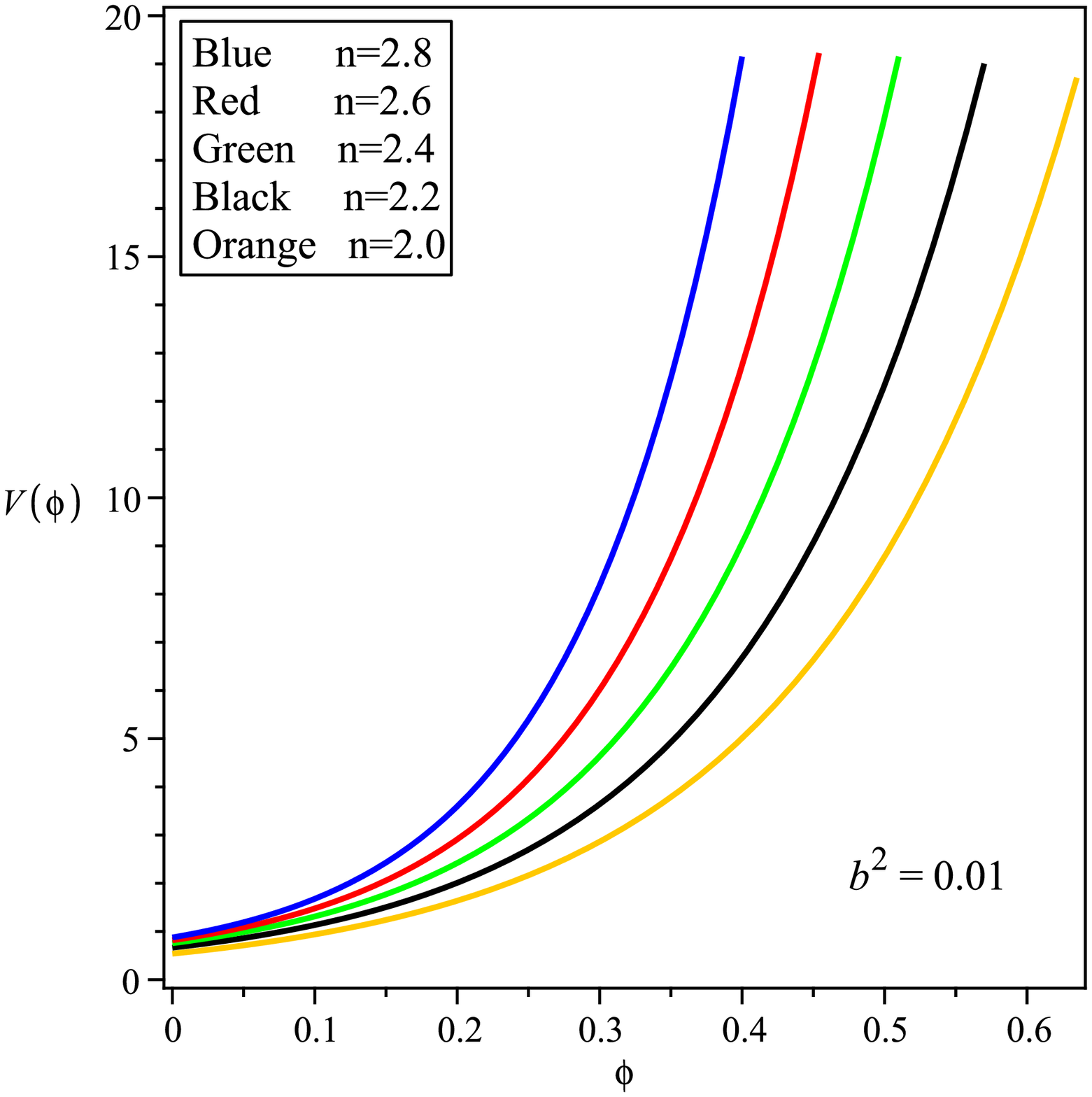}
\caption{The reconstruction of the potential $V(\phi)$ for
original ADE with different model parameter $n$, where $\phi$ is
in unit of $m_p$ and $V(\phi)$ in $\rho_{c0}$. We take here
$\Omega_{m0}=0.28$ and $\Omega_{k}=0.01$. }\label{fig3}
\end{center}
\end{figure}
\begin{figure}[htp]
\begin{center}
\includegraphics[width=8cm]{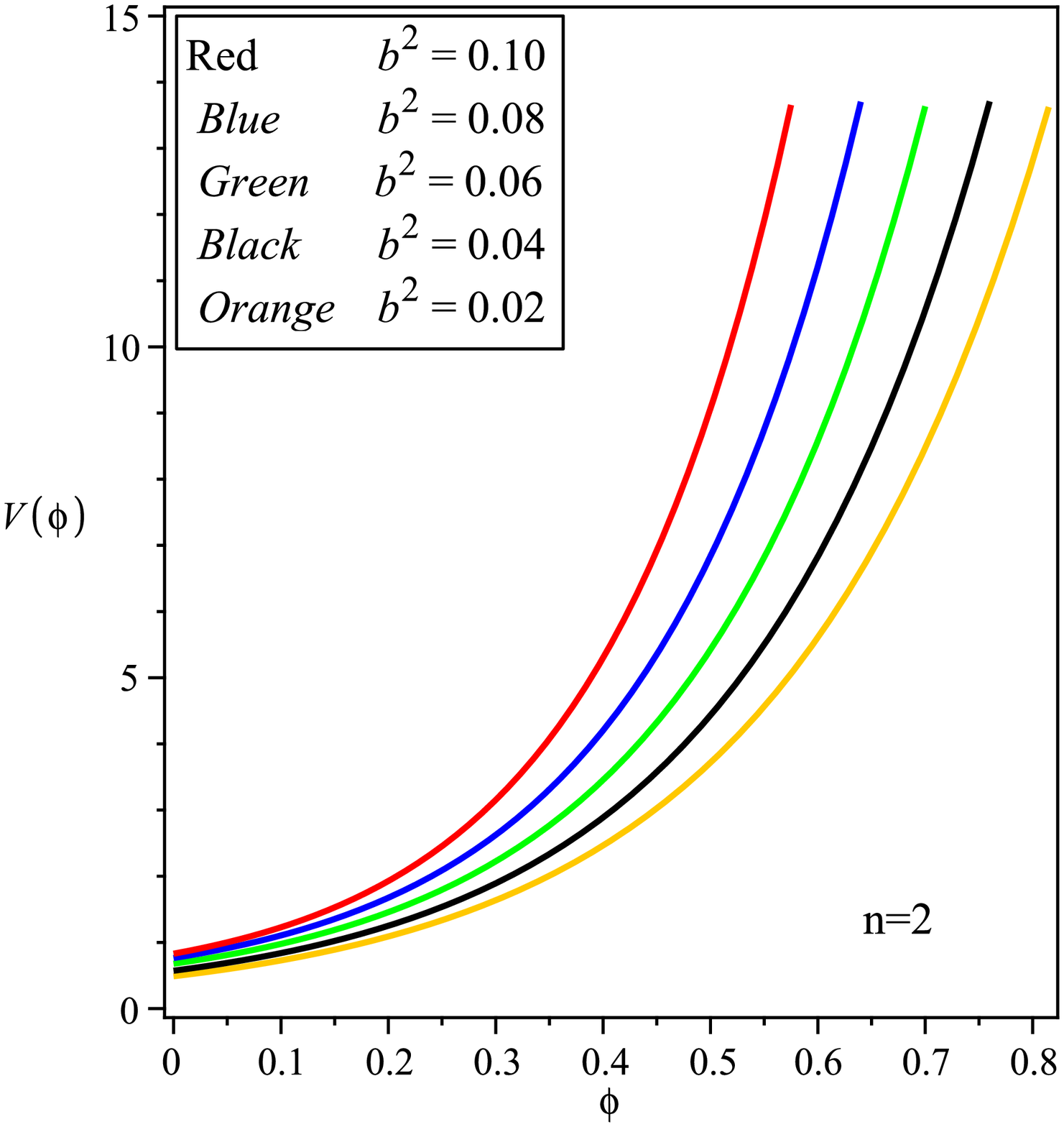}
\caption{The reconstruction of the potential $V(\phi)$ for
original ADE with different interacting parameter $b^2$, where
$\phi$ is in unit of $m_p$ and $V(\phi)$ in $\rho_{c0}$. We take
here $\Omega_{m0}=0.28$ and $\Omega_{k}=0.01$  } \label{fig4}
\end{center}
\end{figure}
\begin{figure}[htp]
\begin{center}
\includegraphics[width=8cm]{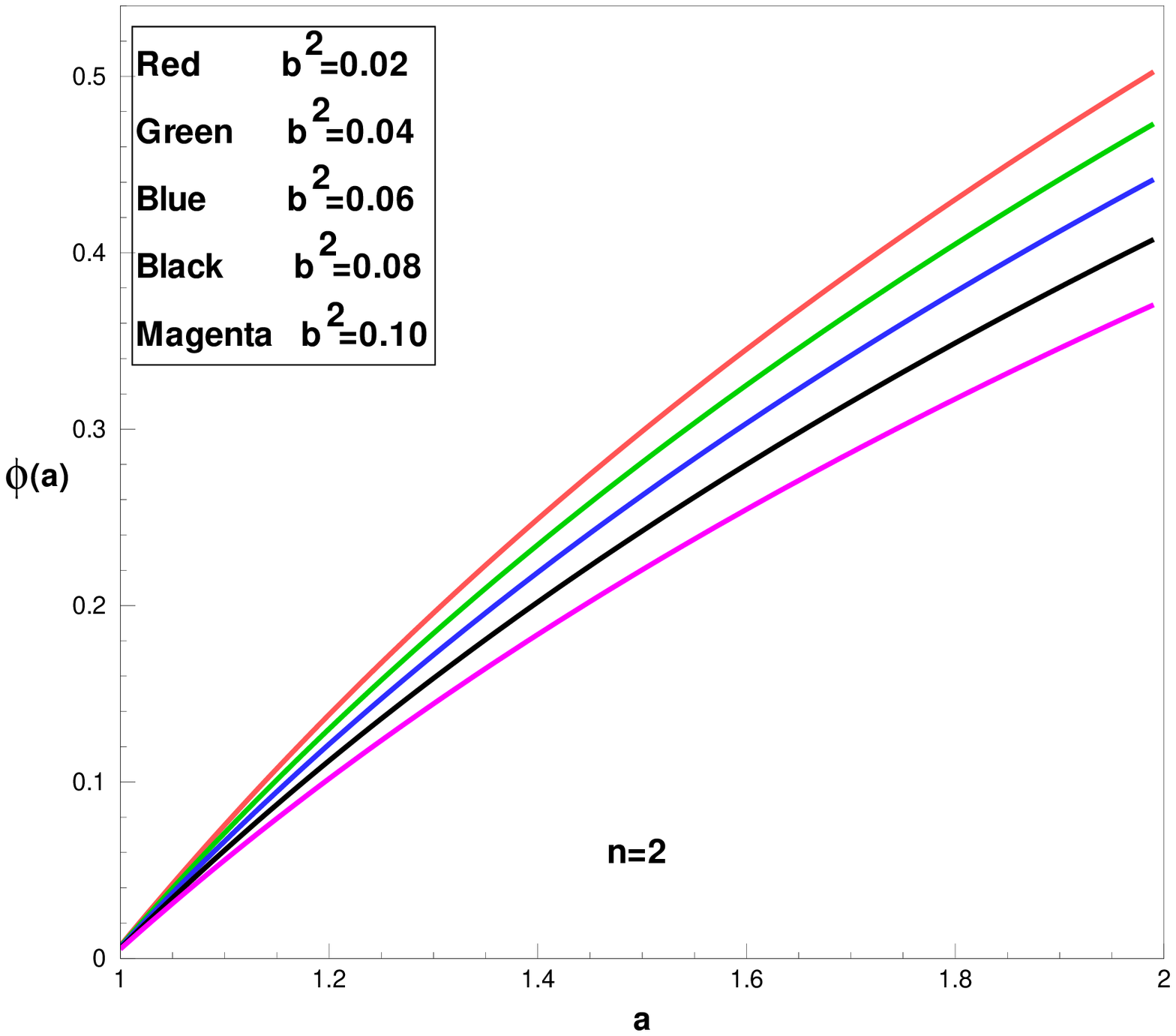}
\caption{The revolution of the scalar-field $\phi(a)$ for
original ADE with different interacting parameter $b^2$, where
$\phi$ is in unit of $m_p$ and we take here $\Omega_{m0}=0.28.$ }
\end{center} \label{fig5}
\end{figure}
\begin{figure}[htp]
\begin{center}
\includegraphics[width=8cm]{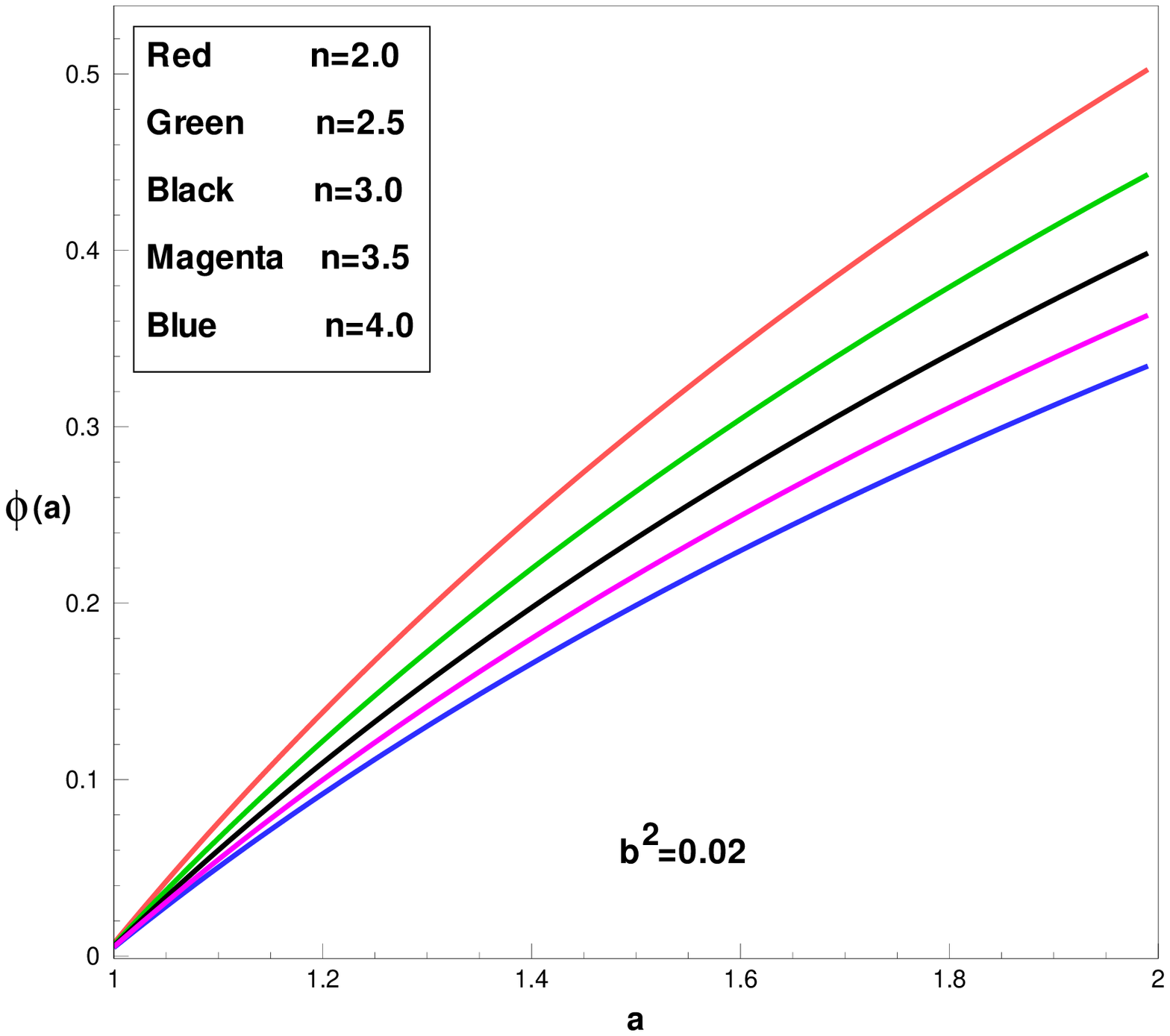}
\caption{The revolution of the scalar-field $\phi(a)$ for
original ADE with different model parameter $n$, where $\phi$ is
in unit of $m_p$ and we take here $\Omega_{m0}=0.28$. }
\label{fig6}
\end{center}
\end{figure}
Differentiating Eq. (\ref{Omegaq}) and using relation
${\dot{\Omega}_D}={\Omega'_D}H$, we reach
\begin{eqnarray}\label{Omegaq2}
{\Omega'_D}=\Omega_D\left(-2\frac{\dot{H}}{H^2}-\frac{2}{n
}\sqrt{\Omega_D}\right),
\end{eqnarray}
where the dot and the prime stand, respectively, for the
derivative with respect to the cosmic time and the derivative with
respect to $x=\ln{a}$. Taking the derivative of both side of the
Friedman equation (\ref{Fried}) with respect to the cosmic time,
and using Eqs. (\ref{Fried2}), (\ref{rho1}), (\ref{Omegaq}) and
(\ref{consm}), it is easy to show that
\begin{eqnarray}\label{Hdot}
\frac{\dot{H}}{H^2}=-\frac{3}{2}(1-\Omega_D)-\frac{\Omega^{3/2}_D}{n}-\frac{\Omega_k}{2}
+\frac{3}{2}b^2(1+\Omega_k).
\end{eqnarray}
Substituting this relation into Eq. (\ref{Omegaq2}), we obtain the
equation of motion of agegraphic dark energy
\begin{eqnarray}\label{Omegaq3}
{\Omega'_D}&=&\Omega_D\left[(1-\Omega_D)\left(3-\frac{2}{n}\sqrt{\Omega_D}\right)
-3b^2(1+\Omega_k)+\Omega_k\right].
\end{eqnarray}
We plot in Figs. 1 and 2 the evolutions of the $w_D$ and
$\Omega_D$ of the original ADE with different interacting
parameter $b^2$. From Fig. 1 we  see that $w_D$ of the agegraphic
dark energy can cross the phantom divide. It was argued
\cite{Wei2} that without interaction ($b^2=0$) $w_D$ is always
larger than $-1$ and cannot cross the phantom divide. In the
presence of the interaction the situation is changed. An
interesting observation from Fig. 1 is that $w_D$ crosses the
phantom divide from $w_D< -1$ to $w_D > -1$. This makes it
distinguishable from many other dark energy models whose $w_D$ can
cross the phantom divide. Fig. 2 shows that at the early time
$\Omega_D\rightarrow0$ while at the late time
$\Omega_D\rightarrow1$, that is the ADE dominates as expected.

Now we suggest a correspondence between the original agegraphic
dark energy and quintessence scalar field namely, we identify
$\rho_\phi$ with $\rho_D$. Using relation
$\rho_\phi=\rho_D={3m_p^2H^2}\Omega_D$ and Eq. (\ref{wq}) we can
rewrite the scalar potential and kinetic energy term as
\begin{eqnarray}\label{vphi2}
V(\phi)&=&m^2_pH^2 \Omega_D\left(3-\frac{\sqrt{\Omega_D}}{n}+\frac{3b^2}{2}\frac{(1+\Omega_k)}{\Omega_D}\right),\\
\dot{\phi}&=&m_pH\left(
\frac{2}{n}{\Omega^{3/2}_D}-3b^2(1+\Omega_k)\right)^{1/2}.\label{dotphi2}
\end{eqnarray}
Using relation $\dot{\phi}=H{\phi'}$, we get
\begin{eqnarray}\label{primephi}
{\phi'}&=&m_p\left(
\frac{2}{n}{\Omega^{3/2}_D}-3b^2(1+\Omega_k)\right)^{1/2}.
\end{eqnarray}
Consequently, we can easily obtain the evolutionary form of the
field by integrating the above equation
\begin{eqnarray}\label{phi}
\phi(a)-\phi(a_0)=\int_{a_0}^{a}{\frac {m_p}{a}\sqrt{
\frac{2}{n}{\Omega^{3/2}_D}-3b^2(1+\Omega_k)}da},
\end{eqnarray}
where $a_0$  is the  present value of the scale factor, and
$\Omega_D$ is given by Eq. (\ref{Omegaq3}). Therefore, we have
established an interacting agegraphic quintessence dark energy
model and reconstructed the potential of the agegraphic
quintessence as well as the dynamics of scalar field.

As one can see from the above equations, the analytical form of
the potential $V=V(\phi)$ is hard to be derived due to the
complexity of the equations, but we can plot the  agegraphic
quintessence potential versus $a$ numerically. For simplicity we
take $\Omega_k\simeq0.01$ fixed in the numerical discussion.
Besides, $\rho_{c0} =3m_p^2 H^2_0$ is the present value of the
critical energy density of the universe. The reconstructed
quintessence potential $V(\phi)$ and the evolutionary form of the
field are plotted in Figs. 3-6, where we have taken
$\phi(a_0=1)=0$. Selected curves are plotted for the different
model parameter $n$ with fixed $b^2$ and different $b^2$ with
fixed $n$, and the present fractional matter density is chosen to
be $\Omega_{m0}=0.28$. From these figures we can see the dynamics
of the potential as well as the scalar field explicitly. They also
show that the reconstructed quintessence potential is steeper in
the early epoch and becomes very flat near today. Consequently,
the scalar field $\phi$ rolls down the potential with the kinetic
energy $\dot{\phi}$ gradually decreasing.
\section{Quintessence reconstruction of NEW ADE  \label{NEW}}
Soon after the original agegraphic dark energy model was
introduced by Cai \cite{Cai1}, a new model of agegraphic dark
energy  was proposed in \cite{Wei2}, while the time scale is
chosen to be the conformal time $\eta$ instead of the age of the
universe. It is worth noting that the Karolyhazy relation
$\delta{t}= \beta t_{p}^{2/3}t^{1/3}$ was derived for Minkowski
spacetime $ds^2 = dt^2-d\mathrm{x^2}$ \cite{Kar1,Maz}. In the case
of the FRW universe, we have $ds^2 = dt^2-a^2d\mathrm{x^2} =
a^2(d\eta^2-d\mathrm{x^2})$.  Thus, it might be more reasonable to
choose the time scale in Eq. (\ref{rho1}) to be the conformal time
$\eta$ since it is the causal time in the Penrose diagram of the
FRW universe. The new agegraphic dark energy  contains some new
features different from the original agegraphic dark energy and
overcome some unsatisfactory points. For instance, the original
agegraphic dark energy suffers from the difficulty to describe the
matter-dominated epoch while the new agegraphic dark energy
resolved this issue \cite{Wei2}. The energy density of the new
agegraphic dark energy can be written
\begin{equation}\label{rho1new}
\rho_{D}= \frac{3n^2 m_{p}^2}{\eta^2},
\end{equation}
where the conformal time $\eta$ is given by
\begin{equation}
\eta=\int{\frac{dt}{a}}=\int_0^a{\frac{da}{Ha^2}}.
\end{equation}
\begin{figure}[htp]
\begin{center}
\includegraphics[width=8cm]{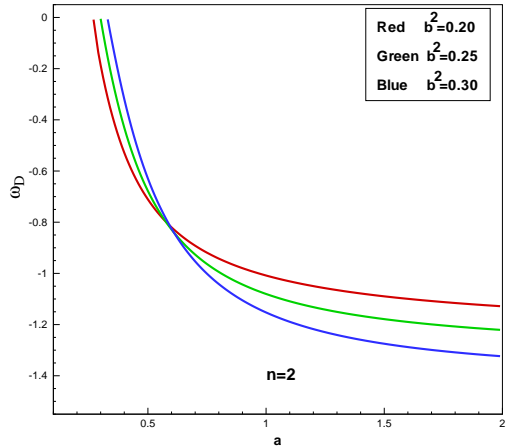}
\caption{The evolution of $w_D$ for new ADE with different
interacting parameter $b^2$. Here we take $\Omega_{D0}=0.72$ and
$\Omega_{k}=0.01.$ }\label{fig7}
\end{center}
\end{figure}
\begin{figure}[htp]
\begin{center}
\includegraphics[width=8cm]{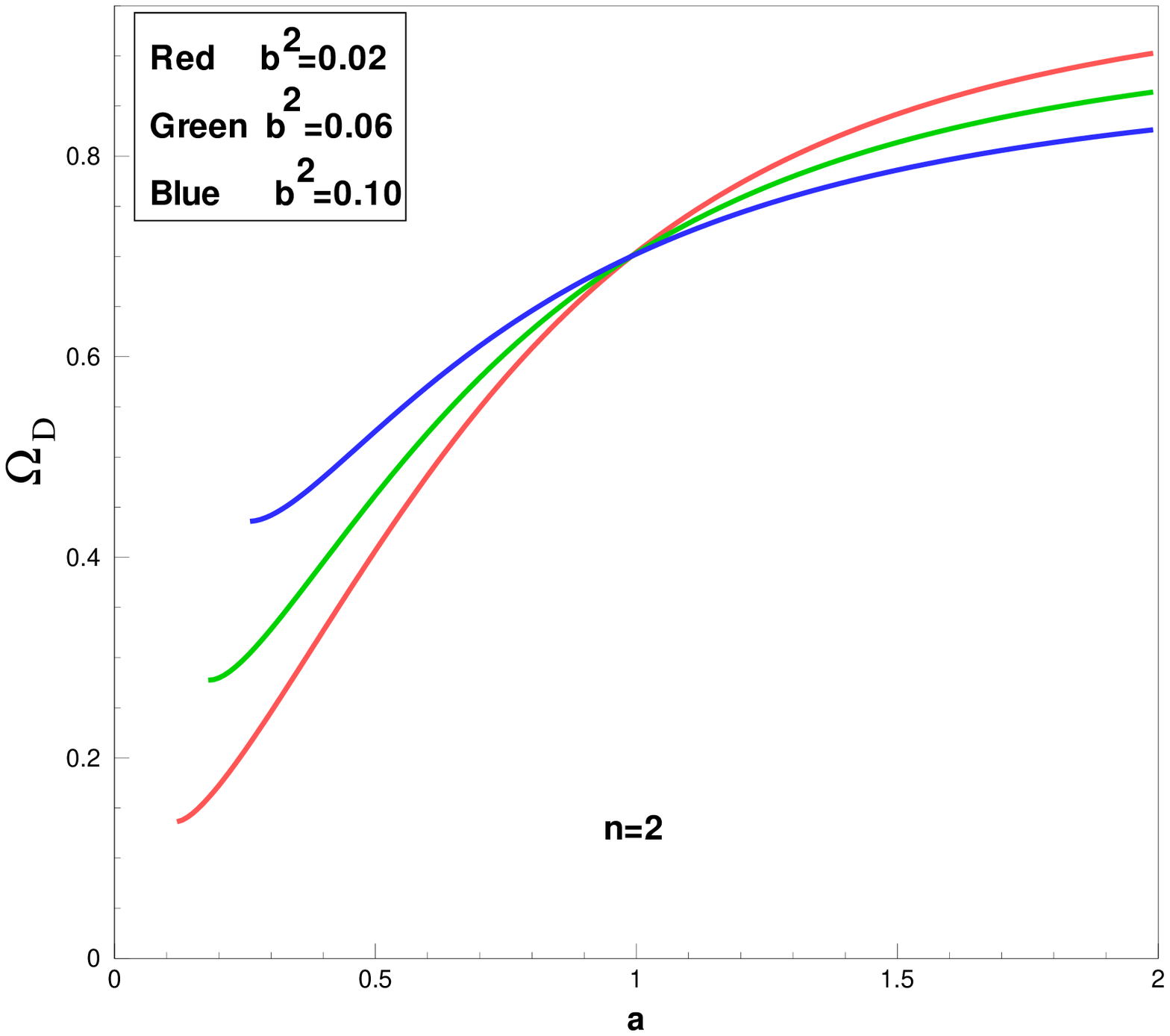}
\caption{The evolution of $\Omega_{D}$ for new ADE with different
interacting parameter $b^2$. Here we take $\Omega_{D0}=0.72$ and
$\Omega_{k}=0.01$.}\label{fig8}
\end{center}
\end{figure}
\begin{figure}[htp]
\begin{center}
\includegraphics[width=8cm]{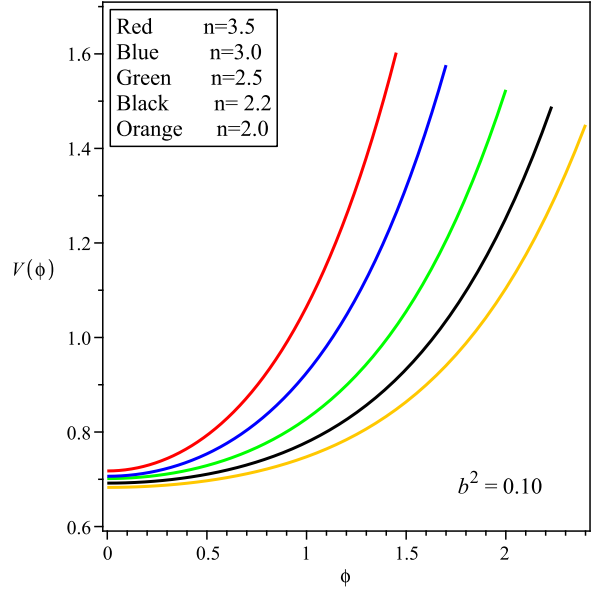}
\caption{The reconstruction of the potential $V(\phi)$ for new ADE
with different model parameter $n$, where $\phi$ is in unit of
$m_p$ and $V(\phi)$ in $\rho_{c0}$. We take here
$\Omega_{m0}=0.28.$ }\label{fig9}
\end{center}
\end{figure}
\begin{figure}[htp]
\begin{center}
\includegraphics[width=8cm]{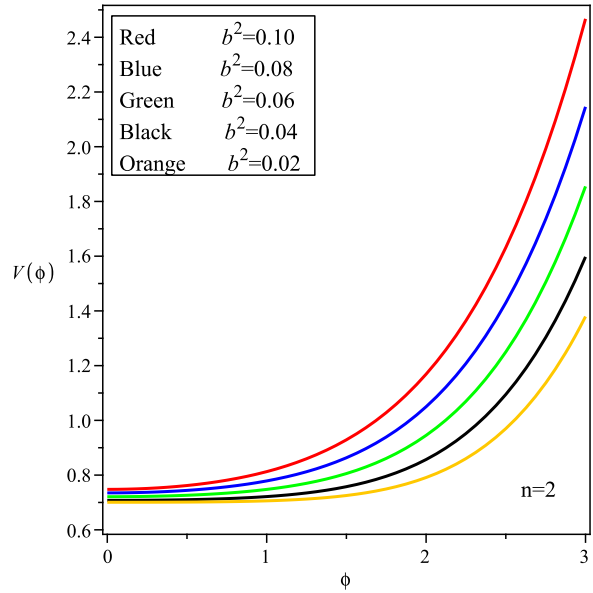}
\caption{The reconstruction of the potential $V(\phi)$ for new ADE
with different interacting parameter $b^2$, where $\phi$ is in
unit of $m_p$ and $V(\phi)$ in $\rho_{c0}$. We take here
$\Omega_{m0}=0.28.$  } \label{fig10}
\end{center}
\end{figure}
\begin{figure}[htp]
\begin{center}
\includegraphics[width=8cm]{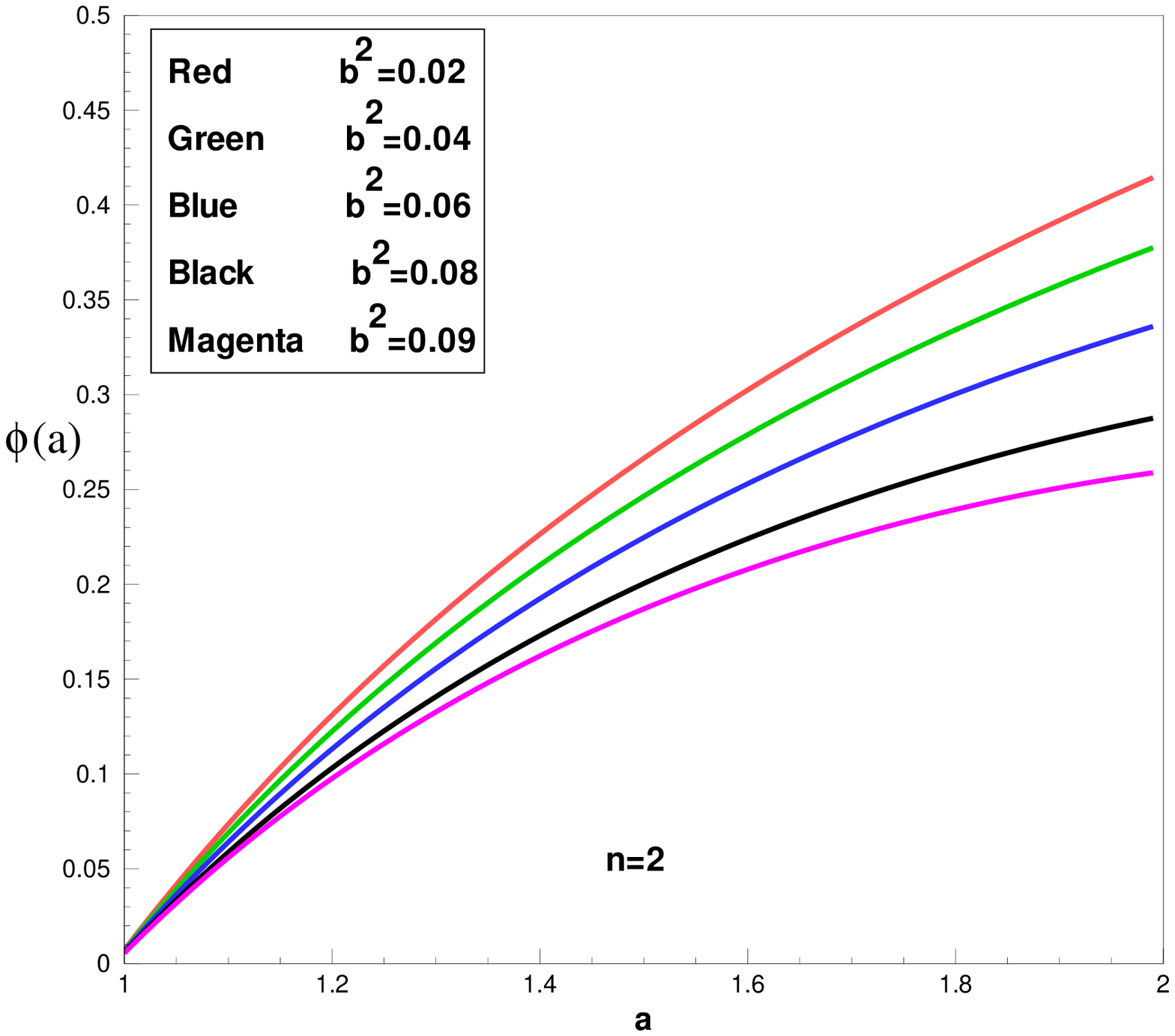}
\caption{The revolution of the scalar-field $\phi(a)$ for new ADE
with different interacting parameter $b^2$, where $\phi$ is in
unit of $m_p$ and we take here $\Omega_{m0}=0.28.$ }
\end{center} \label{fig11}
\end{figure}
\begin{figure}[htp]
\begin{center}
\includegraphics[width=8cm]{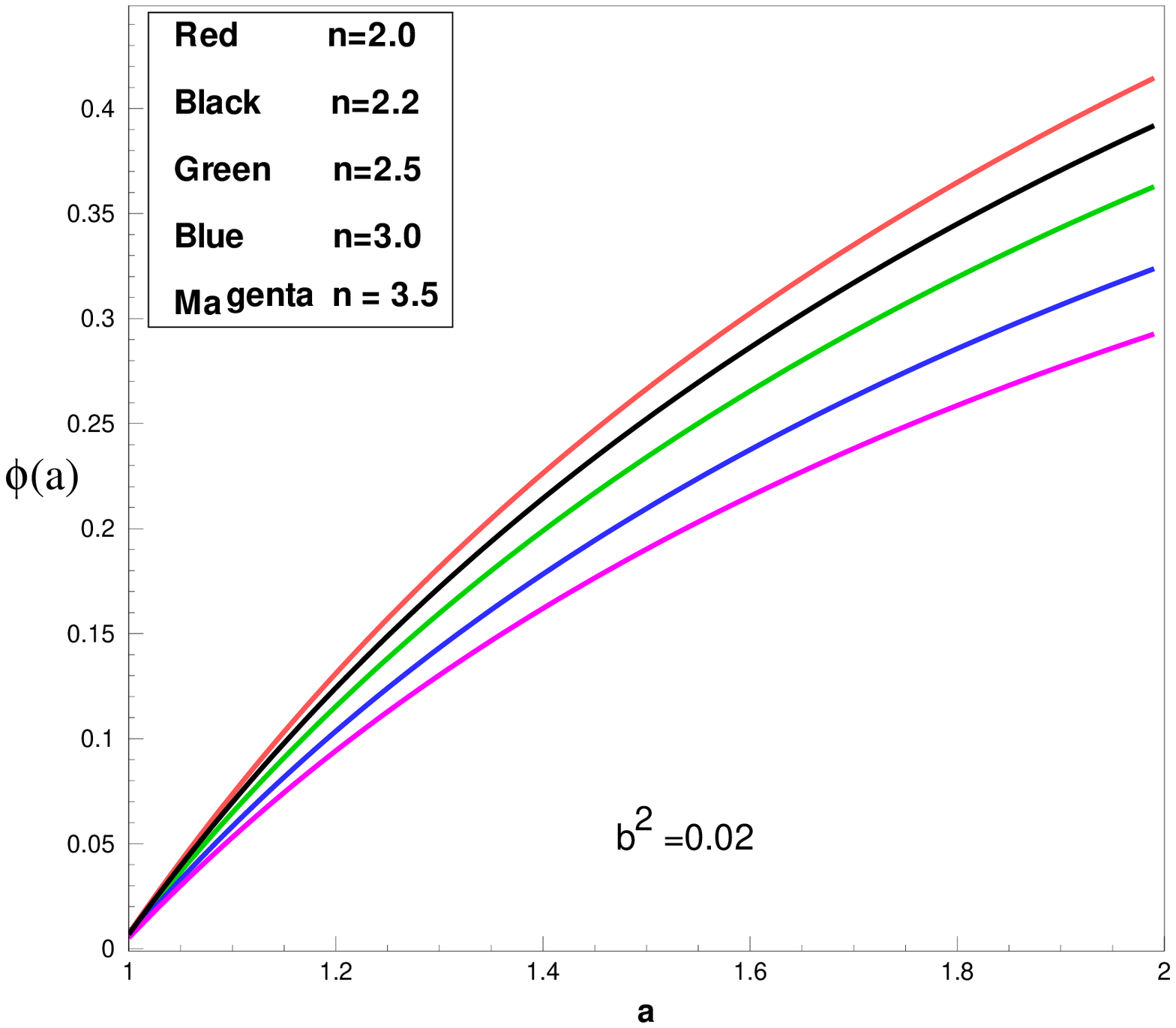}
\caption{The revolution of the scalar-field $\phi(a)$ for new ADE
with different model parameter $n$, where $\phi$ is in unit of
$m_p$ and we take here $\Omega_{m0}=0.28$. } \label{fig12}
\end{center}
\end{figure}
The fractional energy density of the new agegraphic dark energy is
now given by
\begin{eqnarray}\label{Omegaqnew}
\Omega_D=\frac{n^2}{H^2\eta^2}.
\end{eqnarray}
Taking the  derivative with respect to the cosmic time of Eq.
(\ref{rho1new}) and using Eq. (\ref{Omegaqnew}) we get
\begin{eqnarray}\label{rhodotnew}
\dot{\rho}_D=-2H\frac{\sqrt{\Omega_D}}{na}\rho_D.
\end{eqnarray}
Inserting this relation into Eq. (\ref{consq}) we obtain the
equation of state parameter of the new agegraphic dark energy
\begin{eqnarray}\label{wqnew}
w_D=-1+\frac{2}{3na}\sqrt{\Omega_D}-\frac{b^2}{\Omega_D}
(1+\Omega_k).
\end{eqnarray}
The evolution behavior of the new agegraphic dark energy is now
given by
\begin{eqnarray}\label{Omegaq3new}
{\Omega'_D}&=&\Omega_D\left[(1-\Omega_D)\left(3-\frac{2}{na}\sqrt{\Omega_D}\right)
-3b^2(1+\Omega_k)+\Omega_k\right].
\end{eqnarray}
We plot in Figs. 7 and 8 the evolutions of  $w_D$ and $\Omega_D$
of the new ADE with different interacting parameter $b^2$. From
Fig. 7 we  see that $w_D$ of the new ADE can have a transition
from $w_D>-1$ to $w_D < -1$. This is in contrast to the original
ADE model. Fig. 8 indicates that at the late time
$\Omega_D\rightarrow1$, which is similar to the behaviour of the
original ADE.

Next, we reconstruct the new agegraphic quintessence dark energy
model, connecting the quintessence scalar field with the new
agegraphic dark energy. Using Eqs. (\ref{Omegaqnew}) and
(\ref{wqnew}) one can easily show that the scalar potential and
kinetic energy term take the following form
\begin{eqnarray}\label{vphi2new}
V(\phi)&=&m^2_pH^2 \Omega_D\left(3-\frac{\sqrt{\Omega_D}}{na}+\frac{3b^2}{2}\frac{(1+\Omega_k)}{\Omega_D}\right),\\
\dot{\phi}&=&m_pH\left(
\frac{2}{na}{\Omega^{3/2}_D}-3b^2(1+\Omega_k)\right)^{1/2}.\label{dotphi2new}
\end{eqnarray}
Using Eq. (\ref{dotphi2new}), Eq. (\ref{vphi2new}) can be
reexpressed as
\begin{eqnarray}\label{vphiphinew}
V(\phi)&=&3m^2_pH^2 \Omega_D\left(1-\frac{\dot{\phi}^2}{6m^2_pH^2
\Omega_D}\right).
\end{eqnarray}
We can also rewrite Eq. (\ref{dotphi2new}) as
\begin{eqnarray}\label{primephinew}
{\phi'}&=&m_p\left(
\frac{2}{na}{\Omega^{3/2}_D}-3b^2(1+\Omega_k)\right)^{1/2}.
\end{eqnarray}
Therefore the evolution behavior of the scalar field can be
obtained by integrating the above equation
\begin{eqnarray}\label{phinew}
\phi(a)-\phi(a_0)=\int_{a_0}^{a}{\frac {m_p}{a}\sqrt{
\frac{2}{na}{\Omega^{3/2}_D}-3b^2(1+\Omega_k)}da},
\end{eqnarray}
where $\Omega_D$ is now given by Eq. (\ref{Omegaq3new}). In this
way we connect the interacting agegraphic dark energy with a
quintessence scalar field and reconstruct the potential of the
agegraphic quintessence. The reconstructed quintessence potential
$V(\phi)$ and the evolutionary form of the scalar field are
plotted in Figs. 9-12 for different model parameter $n$ with fixed
interacting parameter $b^2$, and different $b^2$ with fixed $n$.
Here the present fractional matter density is chosen to be
$\Omega_{m0}=0.28$.
\section{Conclusions\label{CONC}}
In this paper, we have suggested a correspondence between the
interacting agegraphic dark energy scenarios and the quintessence
scalar field model in a non-flat universe. We have demonstrated
that the agegraphic evolution of the universe can be described
completely by a single quintessence scalar field. We have adopted
the viewpoint that the scalar field models of dark energy are
effective theories of an underlying theory of dark energy. If we
regard the scalar field model as an effective description of such
a theory, we should be capable of using the scalar field model to
mimic the evolving behavior of the interacting agegraphic dark
energy and reconstructing this scalar field model according to the
evolutionary behavior of agegraphic dark energy. We have also
reconstructed the potential of the agegraphic scalar field as well
as the dynamics of the quintessence scalar field which describe
the quintessence cosmology.

\acknowledgments{This work has been supported by Research
Institute for Astronomy and Astrophysics of Maragha, Iran.}

\end{document}